\documentstyle[10lomcon,cite,graphicx]{article}

\def\beq{\begin{equation}}
\def\eeq{\end{equation}}
\def\bea{\begin{eqnarray}}
\def\eea{\end{eqnarray}}

\def\roughly#1{\mathrel{\raise.3ex\hbox
{$#1$\kern-.75em\lower1ex\hbox{$\sim$}}}}

\def\ra{\rightarrow}

\def\journal#1#2#3#4{{\it #1} {\bf #2}, #3 (#4)}
\def\epj{\it Eur. Phys.~J.}
\def\prl{\it Phys. Rev. Lett.}
\def\pl{\it Phys. Lett.}
\def\np{\it Nucl. Phys.}
\def\npps{\it Nucl. Phys. {\bf B} (Proc. Suppl.)}

\def\pr{\it Phys. Rev.}
\def\prp{\it Phys. Rep.}

\def\jp{J. Phys.}
\begin{document}
\begin{flushright}
DESY 01-207 \\
December 2001\\
\end{flushright}

\vspace*{1.5cm}
\begin{center}
{\Large \bf
\centerline{Same-Sign Dilepton Production via Heavy Majorana Neutrinos}
\vskip0.2cm
\centerline{in Proton-Proton Collisions}}
\vspace*{1.5cm}
{\large A.~Ali}
\vskip0.2cm
Deutsches Elektronen-Synchrotron DESY, Hamburg \\
Notkestra\ss e 85, D-22603 Hamburg, FRG\\
\vskip0.5cm   
{\large A.V. Borisov and N.B. Zamorin}
\vskip0.2cm
Physics faculty of Moscow State University,\\
119899 Moscow, Russia

\vspace*{6.0cm}
{\large 
Submitted to the Proceedings of the 10th Lomonosov
Conference on Elementary Particle Physics\\ (Moscow State
University, Moscow, 23 -- 29 August 2001).} 

\end{center}
\thispagestyle{empty}

\newpage
\setcounter{page}{1}
\title{SAME-SIGN DILEPTON PRODUCTION\\ VIA HEAVY
MAJORANA NEUTRINOS\\ IN PROTON--PROTON COLLISIONS}

\author{A.~Ali \footnote{e-mail: ali@x4u2.desy.de}}
\address{Deutsches Elektronen Synchrotron DESY,
Hamburg, Germany}

\author{A.V. Borisov \footnote{e-mail: borisov@ave.phys.msu.su},
N.B.~Zamorin}
\address{Physics Faculty of Moscow State University, 119899 Moscow, Russia}
\vspace*{0.3cm}
\maketitle

\abstracts{~We discuss same-sign dilepton production mediated by
Majorana neutrinos in high-energy proton-proton collisions $pp\ra
\ell^+ \ell^{\prime + }X$ for $\ell,~ \ell^\prime = e,~ \mu,~
\tau$ at the LHC energy $\sqrt{s}=14$ TeV. Assuming one heavy
Majorana neutrino of mass $m_N$, we present discovery limits in
the $\left( m_{N},\left|U_{\ell N}U_{\ell^\prime N}\right|\right)$
plane where $U_{\ell N}$ are the mixing parameters. Taking into
account the present limits from low energy experiments, we show
that at LHC one has sensitivity to heavy Majorana neutrinos up to
a mass $m_{N}\leq 2$ -- $5$ TeV in the dilepton channels $\mu\mu,~
\tau\tau$, and $\mu\tau$, but the dilepton states $e\ell$ will not
be detectable due to the already existing constraints from
neutrinoless double beta decay.}

\section{Introduction}

While impressive, and providing so far the only evidence of new
physics, the solar and atmospheric neutrino experiments do not
probe the nature of the neutrino masses, i.e., they can not
distinguish between the Dirac and Majorana character of the
neutrinos. The nature of neutrino mass is one of the main unsolved
problems in particle physics and there are practically no
experimental clues on this issue\cite{rev}.

If neutrinos are Majorana particles then their mass term violates
lepton number by two units $\Delta L=\pm 2$ \cite{KGP}. Being a
transition between a neutrino and an antineutrino, it can be
viewed equivalently as the annihilation or creation of two
neutrinos. In terms of Feynman diagrams, this involves the
emission (and absorption) of two like-sign $W$-boson pairs
($W^-W^-$ or $W^+ W^+$). If present, it can lead to a large number
of processes violating lepton number by two units, of which
neutrinoless double beta decay ($\beta \beta_{0\nu}$) is a
particular example. The seesaw models \cite{seesaw} provide a
natural framework for generating a small Majorana neutrino mass
which is induced by mixing between an active (light) neutrino and
a very heavy Majorana sterile neutrino of mass $M_N$. The light
state has a naturally small mass $m_\nu \sim m_D^2/M_N \ll m_D$,
where $m_D$ is a quark or charged lepton mass. There is a heavy
Majorana state corresponding to each light (active) neutrino
state. Typical scale for $M_N$ in Grand unified theories (GUT) is
of order the GUT-scale, though in general, there exists a large
number of seesaw models in which both $m_D$ and $M_N$ vary over
many orders of magnitude, with the latter ranging somewhere
between the TeV scale and the GUT-scale \cite{langacker}.

 If $M_N$ is of order GUT-scale, then it is obvious that there are
essentially no low energy effects induced by such a heavy Majorana
neutrino state. However, if $M_N$ is allowed to be much lower, or
if the light (active) neutrinos are Majorana particles, then the
induced  effects of such Majorana neutrinos can be searched for in
a number of rare processes.  Among them neutrinoless double beta
decay, like-sign dilepton states produced in rare meson decays and
in hadron-hadron, lepton-hadron, and lepton-lepton collisions have
been extensively studied. (See, e.g., the papers: $\beta \beta
_{0\nu}$ \cite{moscow-heidelberg,faessler,klapdor,mohapatra}, $K^{+}\ra
\pi^{-}\mu ^{+}\mu ^{+}$ \cite{ls1,zuber1,dib}, $pp\ra\ell^{\pm }
\ell^{\pm }X$ \cite{hsiu}, $pp\ra\ell^{\pm } \ell^{\pm }W^{\mp}X$
\cite{almeida}, $e^{\pm}p\ra \mathop {\nu _{e}} \limits^{\left( {
-} \right)} \ell^\pm \ell^{\prime \pm}X$ \cite{buchm1,flanzetal}.)

Of the current experiments which are sensitive to the Majorana
nature of neutrino, the neutrinoless double beta decay, which
yields an upper limit on the $ee$ element of the Majorana mass
matrix, is already quite stringent \cite{moscow-heidelberg}.
Likewise, precision electroweak physics experiments severely
constrain the mixing elements \cite{buchm2,nar,nar1}.

Taking into account these constraints, we obtain discovery limits
for heavy Majorana neutrinos involved in the process of same-sign
dilepton production in the proton-proton collision:

\beq
pp\ra \ell^+ \ell ^{\prime + }X
\label{col}
\eeq
with $ \ell,~ \ell^\prime = e,~ \mu,~ \tau$ at  the LHC energy
$\sqrt{s}=14~{\rm TeV}$.

\section{Dilepton production in high-energy $pp$ collisions}

We have calculated the cross section for the process (\ref{col})
at high energies,
\beq
\sqrt{s}\gg m_{W},
\label{sw}
\eeq

via an intermediate heavy Majorana neutrino $N$ in the leading effective
vector-boson approximation \cite{evb} neglecting transverse
polarizations of $W$ bosons and quark mixing.  We use the simple
scenario for neutrino mass spectrum
\[
m_{N_{1}}\equiv m_{N}\ll m_{N_{2}}< m_{N_{3}},...,
\]
and single out the contribution of the lightest Majorana neutrino
assuming
\[
\sqrt{s}\ll m_{N_2}.
\]
The cross section for the process in question is then
parameterized by the mass $m_N$ and the corresponding neutrino
mixing parameters $U_{\ell N}$ and $U_{\ell ^{\prime }N}$:

\beq
\sigma \left( pp\ra \ell ^{+}\ell ^{\prime +}X\right)
=\frac{\left( G_{F}m_{W}\right) ^{2}}{8\pi ^{5}}
\left( 1-\frac{1}{2}\delta _{\ell \ell ^{\prime }}\right) \left|
U_{\ell N}U_{\ell ^{\prime }N}\right| ^{2}F\left( E,m_{N}\right)~,
\label{cs}
\eeq
with
\bea
F\left( E,m_{N}\right) =\left( \frac{m_{N}}{m_{W}}\right)
^{2}\int_{z_{0}}^{1}\frac{dz}{z}\int_{z}^{1}\frac{dy}{y}\int_{y}^{1}\frac{dx%
}{x}p\left( x,xs\right) p\left( \frac{y}{x},\frac{y}{x}s\right) \nonumber \\
\times h\left( \frac{z}{y}\right) w\left( \frac{s}{m_{N}^{2}}z\right)~.
\label{F}
\eea
Here, $z_{0}=4m_{W}^{2}/s $, $E=\sqrt{s}$, and
\[
w\left( t\right)= 2+\frac{1}{t+1}-\frac{2\left( 2t+3\right)
}{t\left( t+2\right) }\ln \left( t+1\right)
\]
is the normalized cross section for the subprocess $W^{+}W^{+}\ra
\ell^{+} \ell ^{\prime + }$~ (in the limit (\ref{sw}) it is
obtained from the well-known cross section for $e^{-}e^{-}\ra
W^{-}W^{-}$ \cite{sub} using crossing symmetry). The function
$h(r)$ defined as
\[
h\left( r\right) = -\left( 1+r\right) \ln r-2\left( 1-r\right)
\]
is the normalized luminosity (multiplied by $r$) of $W^{+}W^{+}$
pairs in the two-quark system \cite{evb}, and

\begin{eqnarray*}
p\left( x,Q^{2}\right) = x\sum_{i}q_{i}\left( x,Q^{2}\right)
=x\left( u_{v}+u_{s}+d_{s}+c+b+t\right)
\end{eqnarray*}
is the corresponding quark distribution in the proton.

In the numerical calculation of the cross section (\ref{cs}) the
MRST99 Fortran codes for the parton distributions \cite{mrs} have
been used.

We assume a luminosity $L=100~{\rm fb}^{-1}$ and the mixing
constraints obtained from the precision electroweak data
\cite{nar}
\bea &\sum \left|U_{eN}\right| ^{2}<6.6\times
10^{-3},\quad \sum \left| U_{\mu N}\right| ^{2}<6.0\times 10^{-3}
\left(1.8\times 10^{-3} \right),\nonumber \\ &\sum \left| U_{\tau
N}\right| ^{2}<1.8\times 10^{-2}\left(9.6\times 10^{-3} \right).
\label{etau}
\eea
The bound on the mixing matrix elements involving fermions depends
on the underlying theoretical scenario. There are the single limit
and joint limit \cite{nar,nar1}, obtained by allowing just one
fermion mixing to be present or allowing simultaneous presence of
all types of fermion mixings, respectively. In our analysis, we
have used the conservative constraints for the joint limit case.

We must also include the constraint from the double beta decay
$\beta \beta _{0\nu}$, mentioned above. For heavy neutrinos,
$m_N\gg 1~{\rm GeV}$, the bound is \cite{sub}
\beq
\left|\sum_{N(heavy)} U_{e N} ^{2}\frac{1}{m_{N}}\right|%
<5\times 10^{-5}~{\rm TeV}^{-1}. \label{beta}
\eeq
In calculating the cross sections for the $\ell\tau $ and
$\tau\tau$ processes, we have used the effective value \beq \left|
U_{\tau N}\right| _{eff}^{2}={\rm B}_{\tau\mu }\left| U_{\tau N}\right| ^{2}%
<3.1\times 10^{-3} \label{eff} \eeq with ${\rm B}_{\tau\mu }={\rm
Br}\left( \tau ^{-}\ra \mu ^{-}\overline{\nu }_{\mu }\nu _{\tau
}\right) =0.1737$ \cite{pdg}, as this $\tau$-decay mode is most
suitable for the like-sign dilepton detection at LHC (see, e.g.,
\cite{flanzetal}).

\begin{figure}[htb]
\vspace{-0.4cm}
\begin{minipage}[c]{0.5\textwidth}
%\centering
\includegraphics[scale=0.3]{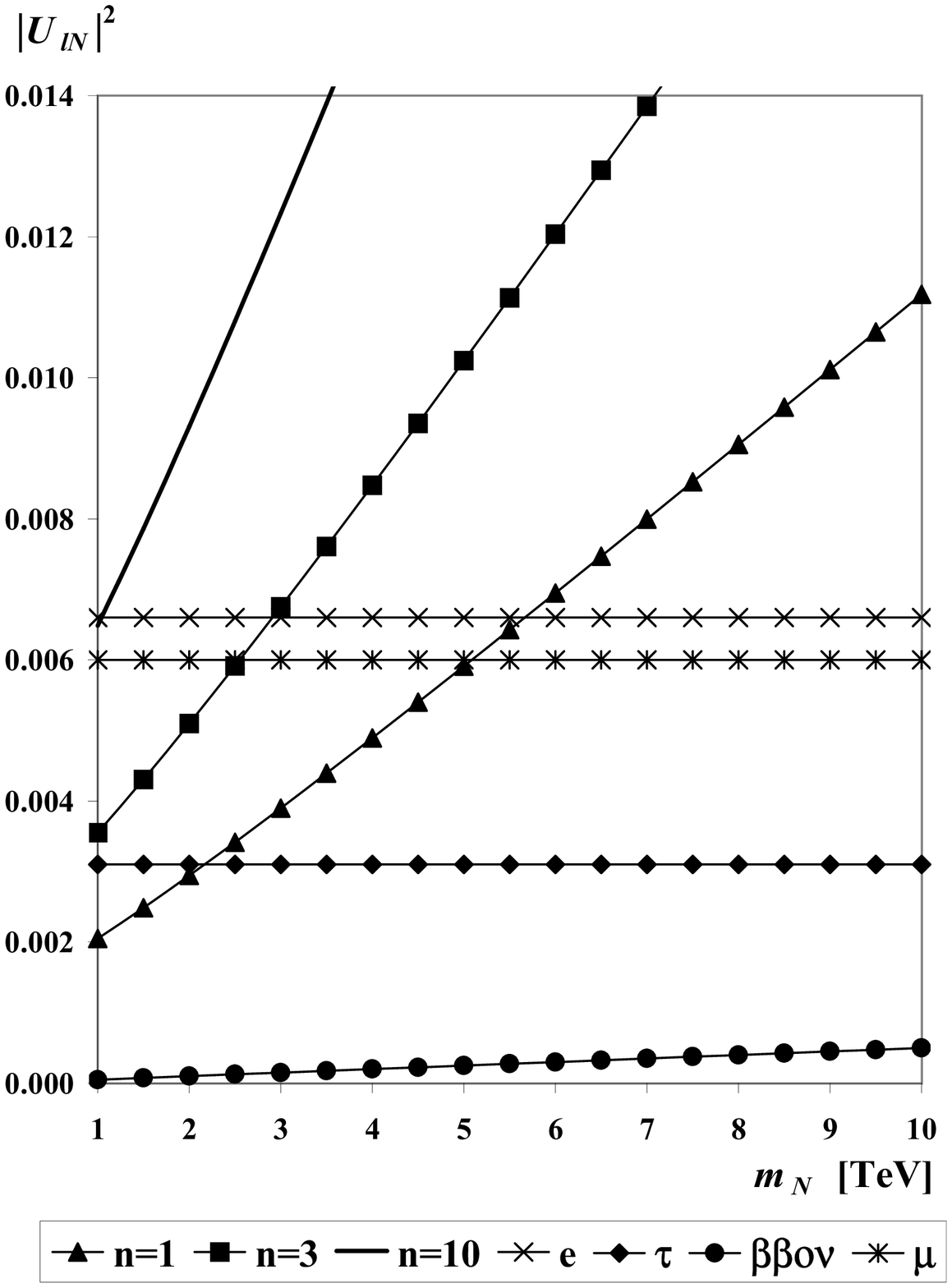}
\end{minipage}
\begin{minipage}[c]{0.49\textwidth}
\vspace{-0.29cm}
%\centering
\includegraphics[scale=0.28]{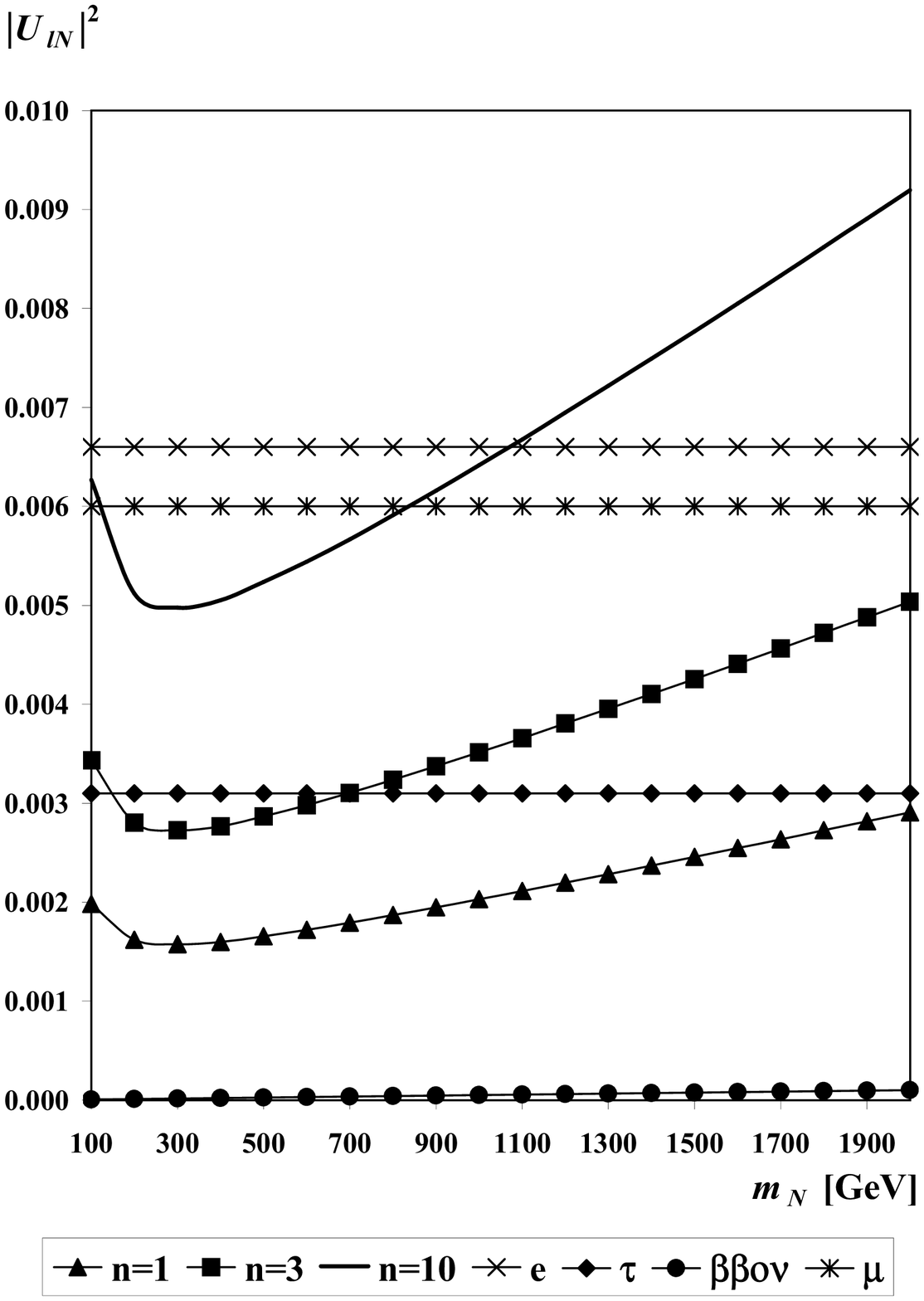}
\end{minipage}
\vspace{-1.4cm}
\caption{{\it Left}: Discovery limits for $pp \to
\ell^+ \ell^+ X$ as functions of $m_{N}$ and $\left|U_{\ell
N}\right|^{2}$ for $E=14~\mbox{TeV}$, $L=100~{\rm fb}^{-1}$ and
various values of $n$, the number of events. We also superimpose
the experimental limit from $\beta \beta _{0\nu}$
(Eq.(\ref{beta})), as well as the experimental limits on
$\left|U_{\ell N}\right|^{2}$ [horizontal lines for $\ell =
e,\,\mu $ (Eq. (\ref{etau})), and $\tau$ (Eq. (\ref{eff}))]. {\it
Right}:~The same as the left figure~but for lighter Majorana
neutrinos.}%\hspace{7cm}\mbox{}%}
\label{Fig2}
\end{figure}

\begin{figure}[htb]
%\vspace{-2cm}
\begin{minipage}[c]{0.5\textwidth}
%\vspace{-1cm}
\includegraphics[scale=0.3]{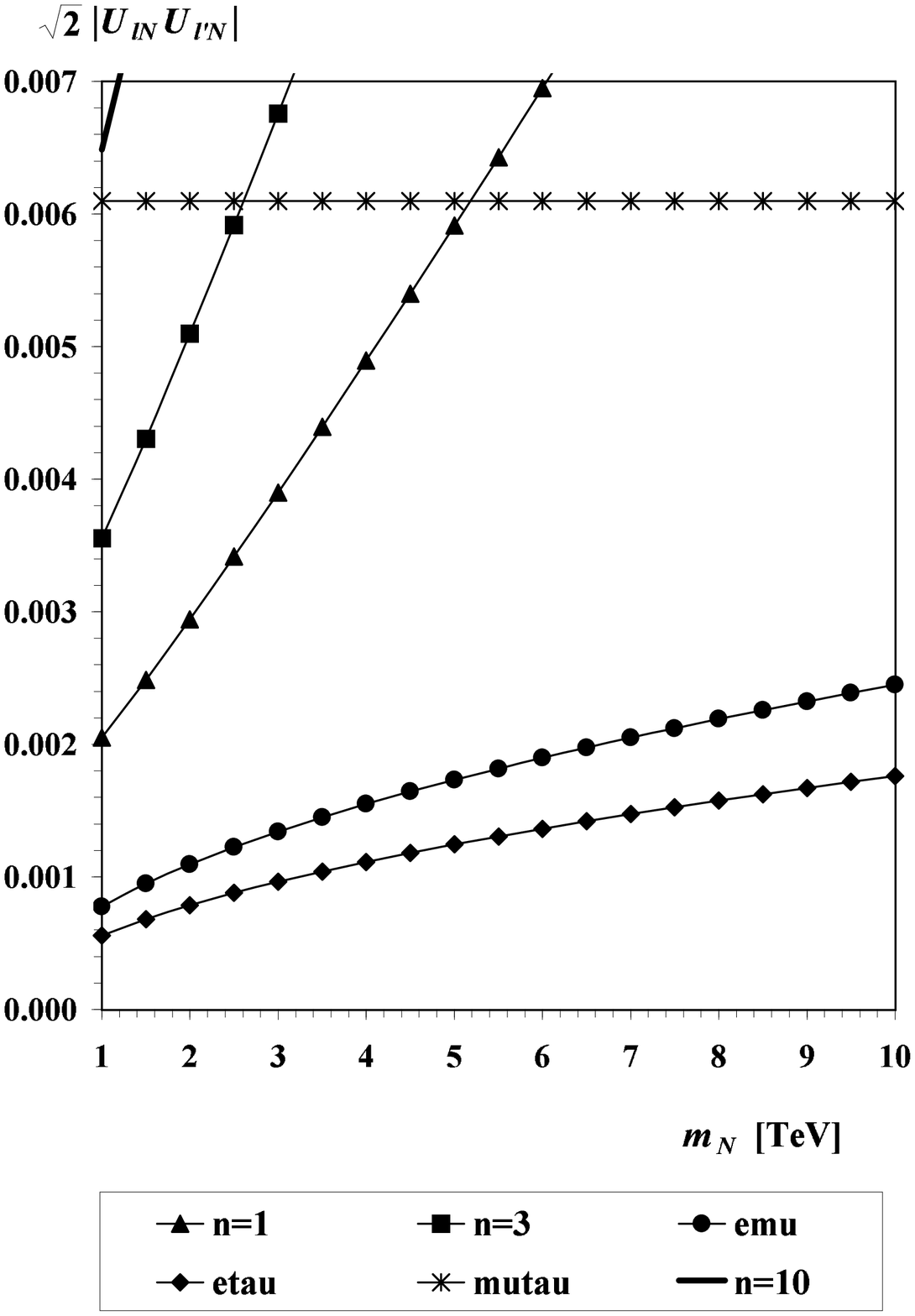}
\end{minipage}
% \hspace*{1cm}
\begin{minipage}[c]{0.49\textwidth}
%\centering
\includegraphics[scale=0.3]{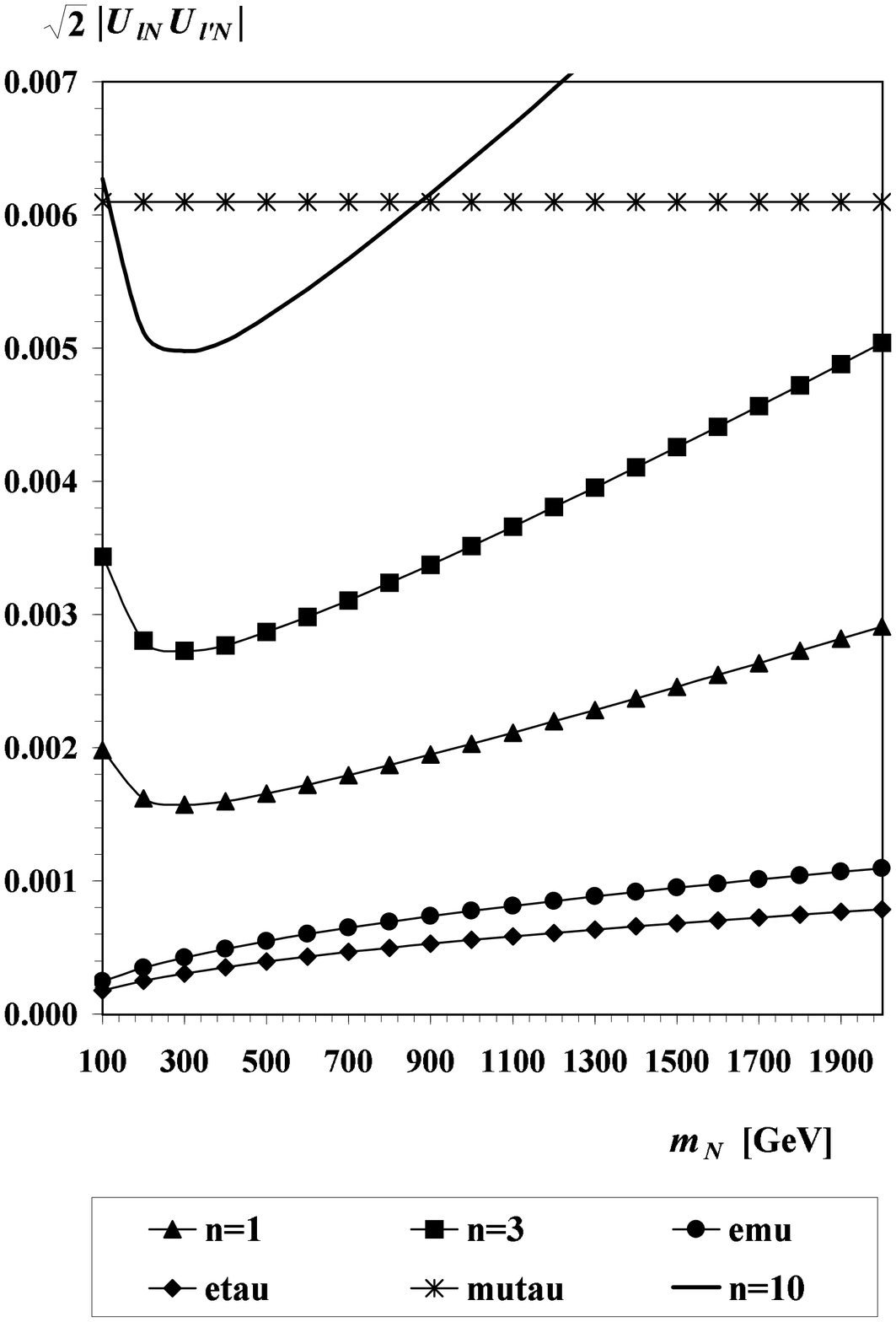}
\end{minipage}
\vspace{-1.2cm}
\caption{{\it Left}: Discovery limits for $pp \to
\ell^+ \ell^{\prime +} X,~\ell \ell ^{\prime }= e\mu ,~e\tau ,~\mu
\tau $. We also superimpose the limits on $\sqrt{2}\left|U_{\ell
N}U_{\ell^{\prime} N}\right|$ obtained from the experimental
limits [Eqs. (\ref{beta}), (\ref{etau}), and~ (\ref{eff})].~{\it
Right}: The same as the left figure but for lighter Majorana
neutrinos.} \label{Fig3}
\end{figure}

Combining the constraints of Eqs. (\ref{etau}), (\ref{eff}), and
(\ref{beta}) and demanding $n = 1,~3,~10$ events for discovery
(i.e.,~ $\sigma L>n$), we present the two-dimensional plot for the
discovery limits in Fig.~\ref{Fig2} for the case of identical
same-sign leptons ($\ell=\ell^\prime$).
Discovery limits for the case of distinct same-sign leptons,~~\mbox{$\ell
\ell ^{\prime }= e\mu ,~e\tau ,~\mu \tau $}, are shown in
Fig.~\ref{Fig3}.

>From Figs.~\ref{Fig2} and \ref{Fig3} we see that  the strong
constraint from $\beta \beta _{0\nu}$  rules out the observation
of the same-sign $e\ell$ processes (with $\ell = e,~ \mu,~ \tau $)
at the LHC. But there are sizable regions of $m_{N}-\left| U_{\ell
N}U_{\ell ^{\prime }N}\right| $ parameter space where observable
signals for the same-sign $\mu\mu,~\tau\tau,$ and $\mu\tau$
processes mediated by heavy Majorana neutrinos of mass $m_N \leq
2$ -- $5~{\rm TeV}$ can be expected. Hence, LHC experiments have a
sensitivity to the matrix elements of the Majorana mass matrix in
the second and third rows of this matrix.

We have also worked out a large number of rare meson decays of the
type $M^{+}\ra M^{\prime -}\ell ^{+}\ell ^{\prime+}$, both for the
light and heavy Majorana neutrino scenarios, and argued that the
present experimental bounds on the branching ratios are too weak
to set reasonable limits on the effective Majorana masses (for
details, see \cite{abz}).

\section{Conclusion}

In conclusion, same-sign dilepton production at LHC will provide
non-trivial constraints on the Majorana mass matrix in the $\mu
\mu$, $\mu \tau$ and $\tau \tau$ sector.

\section*{Acknowledgments}

We thank Christoph Greub, David London and Enrico Nardi for
helpful discussions and communication. A.V.B. thanks DESY for its
hospitality and partial support.


\section*{References}
\begin{thebibliography}{99}

\bibitem{rev}
For a review, see  K. Zuber, \journal{\prp}{305}{295}{1998};
B. Kayser, preprint hep-ph/0104147; S.M. Bilenky, preprint hep-ph/0110306.

\bibitem{KGP}
For a nice introduction to the physics of Majorana neutrinos, see
B. Kayser, F. Gibrat-Debu,  F. Perrier, {\it ``The Physics of the
Massive Neutrinos"} (World Scientific, Singapore) 1989.

\bibitem{seesaw}
M. Gell-Mann, P. Ramond, R. Slansky, in {\it ``Supergravity"},
eds. D. Freedman, P. van Nieuwenhuizen (North Holland,
Amsterdam)~315, 1979;\\ T. Yanagida, in {\it Proceedings of the
Workshop on Unified Theory and Baryon Number in the Universe},
eds. O. Sawada, A. Sugamoto (KEK, Tsukuba, Japan) 1979; R.N. Mohapatra
and G. Senjanovic, \journal{\prl}{44}{912}{1980}.

\bibitem{langacker}
See, for example, P. Langacker, \journal{\npps}{100}{383}{2001}.

\bibitem{moscow-heidelberg} L. Baudis et al.,
\journal{\prl}{83}{411}{1999}.

\bibitem{faessler} A. Faessler,  F. Simkovic,
\journal{\jp}{G~24}{2139}{1998}.

\bibitem{klapdor}
H.V. Klapdor-Kleingrothaus, \journal{\it Springer Tracts in Modern
Physics}{163}{69}{2000}; preprint hep-ph/0102277.

\bibitem{mohapatra}
R.N. Mohapatra, \journal{\pr}{D~34}{909}{1986}.

\bibitem{ls1}
L.S. Littenberg,  R.E.~Shrock, \journal{\pl}{B~491}{285}{2000}.

\bibitem{zuber1} K. Zuber, \journal{\pl}{B~479}{33}{2000}.

\bibitem{dib} C. Dib, V. Gribanov, S. Kovalenko,  I. Schmidt,
\journal{\pl}{B~493}{82}{2000}.

\bibitem{hsiu} H. Tso-hsiu, C. Cheng-rui,  T. Zhi-jian,
\journal{\pr}{D~42}{2265}{1990}; A.~Datta, M. Guchait,  D.P. Roy,
\journal{\pr}{D~47}{961}{1993}; A. Ferrari et~al.,
\journal{\pr}{D~62}{013001}{2000}.

\bibitem{almeida} F.M.L. Almeida Jr, Y.A. Coutinho, J.A. Martins Sim\~oes,
 M.A.B. do Vale, \journal{\pr}{D~62}{075004}{2000}.

\bibitem{buchm1}
W. Buchm\"uller, C. Greub, \journal{\np}{B~363}{345}{1991}.

\bibitem{flanzetal} M. Flanz, W. Rodejohann, K. Zuber,
\journal{\pl}{B~473} {324}{2000} (Erratum,
\journal{\pl}{B~480}{418}{2000}).

\bibitem{buchm2}
W. Buchm\"uller, C. Greub,  H.-G. Kohrs, \journal{\np}{B
370}{3}{1992}.

\bibitem{nar} E. Nardi, E. Roulet, D. Tommasini, \journal{\pl}{B 344}
{225}{1995}; E.~Nardi, private communication.

\bibitem{nar1} E. Nardi, E. Roulet, D. Tommasini, \journal{\pl}{B 327}
{319}{1994}.

\bibitem{evb} S. Dawson, \journal{\np}{B 249}{42}{1985}; M.
Chanowitz,  M.K. Gaillard, \journal{\pl}{B 142}{85}{1984}; G.L.
Kane, W.W. Repko, W.B. Rolnik, \journal{\pl}{B 148}{367}{1984}; I.
Kuss, H. Spiesberger, \journal{\pr}{D 53}{6078}{1996}.

\bibitem{sub} G. B\'elanger, F. Boudjema, D. London,  H. Nadeau,
\journal{\pr} {D~53}{6292}{1996}; D. London, preprint
hep-ph/9907419.

\bibitem{mrs} A.D. Martin, R.G. Roberts, W.J. Stirling, R.S.
Thorne, \journal{\epj}{C 14}{133}{2000}.

\bibitem{pdg} D.E. Groom et al. (PDG Collaboration),
\journal{\epj}{C~15}{1}{2000}.

\bibitem{abz} A. Ali, A.V. Borisov, N.B. Zamorin, \journal{\epj}{C~21} {123}{2001}.

\end{thebibliography}
\end{document}